# Design and Performance Evaluation of an Optimized Disk Scheduling Algorithm (ODSA)

Sourav Kumar Bhoi
Student, M.Tech
Department of CSE
NIT, Rourkela

Sanjaya Kumar Panda
Student, M.Tech
Department of CSE
NIT, Rourkela

Imran Hossain Faruk
Student, M.Tech
Department of CSE
NIT, Rourkela

## ABSTRACT
Management of disk scheduling is a very important aspect of operating system. Performance of the disk scheduling completely depends on how efficient is the scheduling algorithm to allocate services to the request in a better manner. Many algorithms (FIFO, SSTF, SCAN, C-SCAN, LOOK, etc.) are developed in the recent years in order to optimize the system disk I/O performance. By reducing the average seek time and transfer time, we can improve the performance of disk I/O operation. In our proposed algorithm, Optimize Disk Scheduling Algorithm (ODSA) is taking less average seek time and transfer time as compare to other disk scheduling algorithms (FIFO, SSTF, SCAN, C-SCAN, LOOK, etc.), which enhances the efficiency of the disk performance in a better manner.

## 1. INTRODUCTION
In multiprogrammed operating systems, many processes may be generating requests for reading and writing disk records. These processes sometimes make requests faster than they can be serviced by the moving-head disks, waiting lines or queues build up for each device [1]. Some of the computing systems work on *First Come First Serve (FCFS)* technique in which the request coming first is served first. Disk scheduling technique is a process of allocating services to the requests in well manner. It reduces the effect of starvation of the requests which degrade the performance of the disk scheduling process. There are many disk scheduling algorithms such as *FCFS, SSTF, SCAN, C-SCAN, LOOK* etc. which helps in reducing the average seek time. The main aim of disk scheduling algorithms is to reduce or minimize the seek time for a set of requests. The disk performance can be optimized by installing a hard disk that can result in high transfer rates. Hard disk is a collection of platters. We store information by recording it magnetically on the platters. A read/write head is located above the platter. The space of the platter is logically divided into tracks. The tracks are then subdivided into sectors. The set of tracks that are at one arm position forms a cylinder. The heads are attached to a disk arm, which all the heads as a unit. Disks are currently four orders of magnitude slower than main memory, so to increase the performance many researches are going on to enhance the efficiency of disks [2]. By reducing the average seek time we can improve the performance of disk I/O operation. In our proposed algorithm, optimize disk scheduling algorithm (ODSA) is taking less average seek time and transfer time as compare to other disk scheduling algorithms (FIFO, SSTF, SCAN, C-SCAN, LOOK, etc.), which enhances the scheduling of disk I/O requests in a better manner.

### 1.1 Disk Performance Parameters
The disk I/O operations mainly depend on the computer system, the operating system, and the nature of the I/O channel and disk controller hardware [10]. The time taken to position the head at the desired track is called *Seek Time*. The time taken to reach the desired sector is called *Latency Time* or *Rotational Delay*. The sum of seek time and rotational delay is called *Access Time*. The *Transfer Time* mainly depends on the rotational speed of the disk. The total number of bytes transferred, divided by the total time between the first request for service and the completion of the last transfer is called *Bandwidth* [3]. These are some of the disk performance parameters to enhance the efficiency of the disk by which we can improvise or optimize the scheduling.

### 1.2 Disk Scheduling Algorithms
Disk scheduling algorithms are the algorithms to allocate the services to the requests [11]. There are many disk scheduling algorithms such as *FCFS, SSTF, SCAN, C-SCAN, and LOOK* etc. which helps in scheduling the requests. *First Come First Serve (FCFS)* serves the request coming first. But it does not provide the fastest service. It is simple to implement. The average head movement in the algorithm is too high. *Shortest Seek Time Next (SSTF)* selects the request with minimum seek time from the current head position. It gives substantial improvement in comparison to FCFS. *Scan* algorithm is called elevator algorithm. In this the disk arm moves from one end of the disk and move towards other end, while in mean time all requests are servicing until it gets other end of the disk. Comparing with FCFS and SSTF it gives better performance. *C-Scan* scheduling algorithm is called *Circular scan*. The head moves from one end to other end of the disk, servicing the request along the way. The waiting time increases in the algorithm. *Look* scheduling the disk arm moves across the full width of the disk. The arm goes as far as the final request in each direction and reverses immediately. So these are some of the disk scheduling algorithms to serve the requests.

### 1.3 Related Work Done
In the recent years many researches has been done for enhancing the disk performance. Z. Dimitrijevic, R. Rangaswami and E. Y. Chang have presented Semi-preemptible I/O, which divides disk I/O requests into small temporal units of disk commands to improve the preemptibility of disk access [4]. Cheng - Han Tsai, Tai - Yi Huang, Edward T. - H. Chu, Chun-Hang Wei and Yu - Che Tsai propose a novel real-time disk-scheduling algorithm called WRR - SCAN (Weighted-Round-Robin-SCAN) to provide quality guarantees for all in-service streams encoded at variable bit rates and bounded response times for aperiodic jobs [5]. Daniel L. Martens and Michael J. Katchabaw developed a new disk scheduling algorithm focuses on dynamic scheduling algorithm selection and tuning [6]. Worthington and Ganger examined theimpact of complex logical-to-physical mappings and large prefetching caches on scheduling effectiveness [7]. Muqaddas and Abdulsalam made a simulator (Disksims) [8]. Burkhard and Palmer reduced the required flashmemory by a factor of more than thirty therebyreducing the manufacturing cost per drive [9].

## 2. ODSA ALGORITHM

The main aim of our proposed ODSA algorithm is to improve the disk performance by reducing average seek time of the disk scheduling algorithm. So that there will be faster data transfer. The main goal behind all is to enhance the system performance.

### 2.1 Proposed ODSA Algorithm

In our proposed ODSA algorithm, the requests in the disk queue are to be sorted according to the track number requested. Then we calculate the absolute difference between the initial disk head position (IDHP) and the lowest track request (LTR) of disk queue and absolute difference of the initial disk head position (IDHP) and the highest track request (HTR)of the disk queue. If (|IDHP – LTR|) is greater than (|IDHP – HTR|), then we scan the requests in ascending order starting from the initial position and if (|IDHP – LTR|) is less than (|IDHP – HTR|), then scanning starts in descending order (Highest track number to lowest track number). If (|IDHP – LTR|) is equal to (|IDHP – HTR|), then scanning can start from any of the side. Finally, we calculate the average seek time and the transfer time. We calculate the transfer time by using the following formula shown in equation 1:

$T_a = T_s + ( 1/2R ) + ( B/RN )$       (1)

$T_a$ = Transfer Time
$T_s$ = Average Seek Time
B = Number of bytes to be transferred
N = Number of bytes on track
R = Rotation speed in revolutions per second

The pseudocode of the algorithm is represented in figure 1 and figure 2 represents the flowchart of the algorithm.

1. if (DQ != NULL)
    //DQ = a Disk Queue with requests for accessing tracks
2. Read IDHP
    //IDHP = Initial Disk Head Position
3. All the TRs present in DQ are sorted in ascending order (numerical order)
    //TR = Track Request
    // n = number of TRs
4. if (| IDHP - LTR |) < (| IDHP - HTR |){
        IST = | IDHP - LTR |
        Scanning will start from the LTR}
   else if (| IDHP - LTR |) > (| IDHP - HTR |) {
        IST = | IDHP - HTR |
        Scanning will start from the HTR}
   else    {
        Scanning can be done from any of the end
        IST = | IDHP - LTR | or| IDHP -    HTR |
        }
   end if
   //LTR = Lowest Track Request
   //HTR = Highest Track Request
   //IST = Initial Seek Time
5. ST ← 0
   // initializing ST to 0
   //ST = Seek Time
6. for i = 1 to n
        ST = ST + | $TR_{i+1}$ - $TR_i$ |
   end for
   //i = loop variable
7. ST = ST + IST
8. Calculate AST and TT
   //AST = Average Seek Time
   //TT = Transfer Time
   end if

**Fig 1: Pseudocode of ODSA Algorithm**

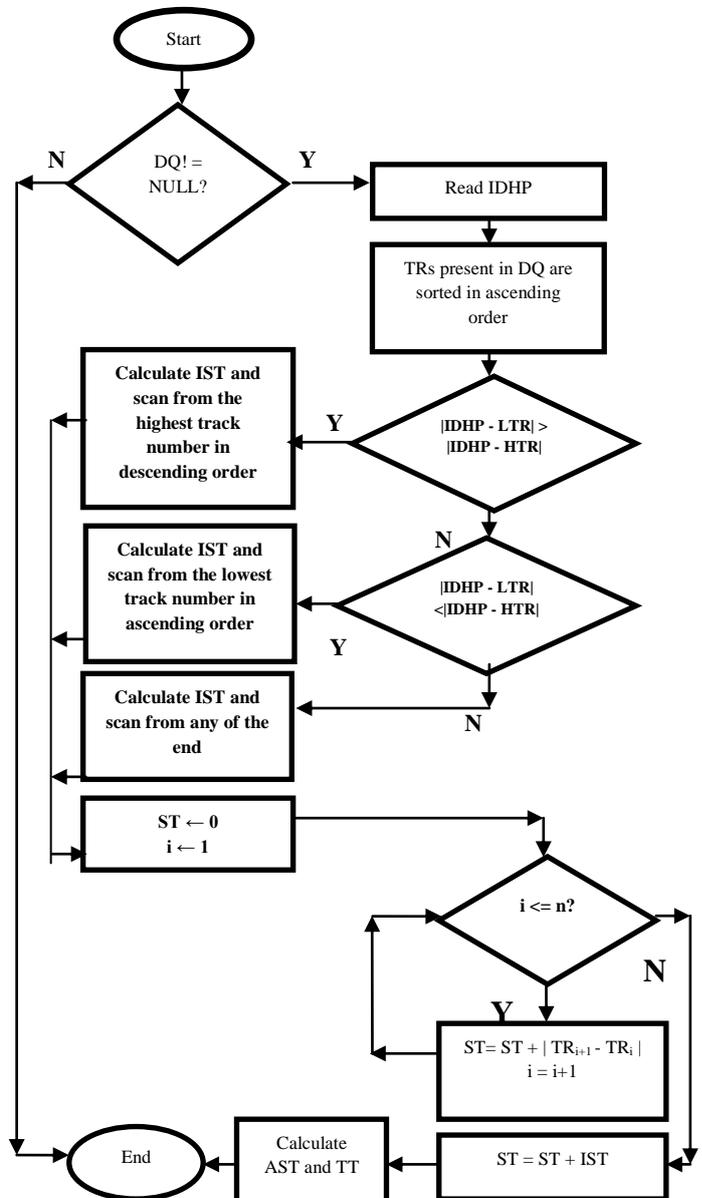

**Fig 2: Flowchart of ODSA Algorithm**

## 2.2 Illustration

We have the following requests to serve (25, 10, 151, 170, 62, 46, 74 and 111). Initially the disk header is at 45 and the minimum track number is 0 and maximum track number taken is 180. Then according to the algorithm we first sort the requests as (10, 25, 46, 62, 74, 111, 151 and 170). Then we calculate (| IDHP – LTR |) as 35 and (| IDHP – HTR |) as 125. So (| IDHP – LTR |) is smaller than (| IDHP – HTR |), then we scan in ascending order starting from 45 then 10 and so on. After calculating the average seeks time 24.375 from the algorithm we calculate the transfer time as 24.38691.

## 3. PERFORMANCE EVALUATION

### 3.1 Assumptions Taken

All requests are independent of each other and have equal priority. The requests are initially stored in the request queue. All the cases taken are ideal in nature. The disk storage capacity is measured in Gigabytes. We have taken a disk of 400 Gigabytes containing sector per tracks = 63, sector size = 512 bytes, cylinders = 16,383, total sectors = 781,422,768, N = 32,256 bytes, B = 30,000 bytes and R = 120 rps. By using these data we calculate the transfer time. The minimum track number is 0 and maximum track number taken is 180.

### 3.2 Performance Parameters

The performance parameters we have taken for experimental analysis is as follows:

1) *Seek Time* (ST): The average seek time should be less for better performance.
2) *Transfer Time* (TT): The transfer time should be less for faster accessing of data.

### 3.3 Experiments Performed

To evaluate the performance of our proposed algorithm, we have taken three different cases. In each case, we have compared the experimental results of our proposed algorithm with other disk scheduling algorithms.

**Case 1:** We have taken the following track requests for accessing the tracks as (25, 10, 151, 170, 62, 46, 74 and 111) and the initial disk head position is at 45. Table 1 shows the comparison of all the algorithms with our proposed algorithm. Figure 3, Figure 4, Figure 5, Figure 6, Figure 7 and Figure 8 shows the representation of FIFO, SSTF, SCAN, C-SCAN, LOOK and ODSA respectively. Figure 9 and Figure 10 shows the comparison of average seek time and transfer time respectively.

**Table 1: Comparison of all algorithms with ODSA**

| Algorithms | Average Seek Time | Transfer Time |
|---|---|---|
| FIFO | 48 | 48.01191 |
| SSTF | 35.625 | 35.63691 |
| SCAN | 38.125 | 38.13691 |
| C-SCAN | 42.5 | 42.51191 |
| LOOK | 37.5 | 37.51191 |
| ODSA | 24.375 | 24.38691 |

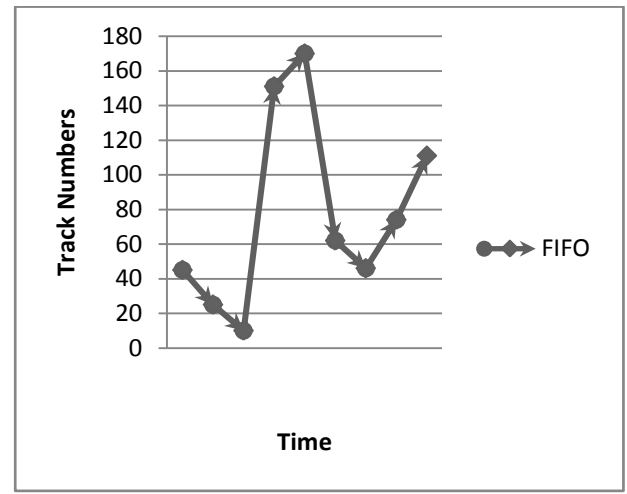

**Fig 3: Representation of FIFO (Case 1)**

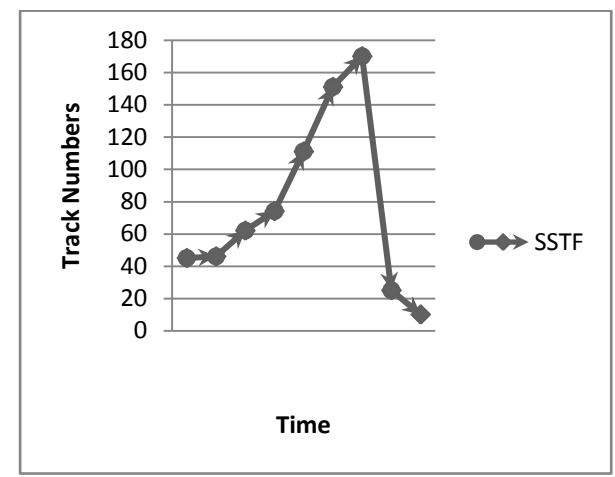

**Fig 4: Representation of SSTF (Case 1)**

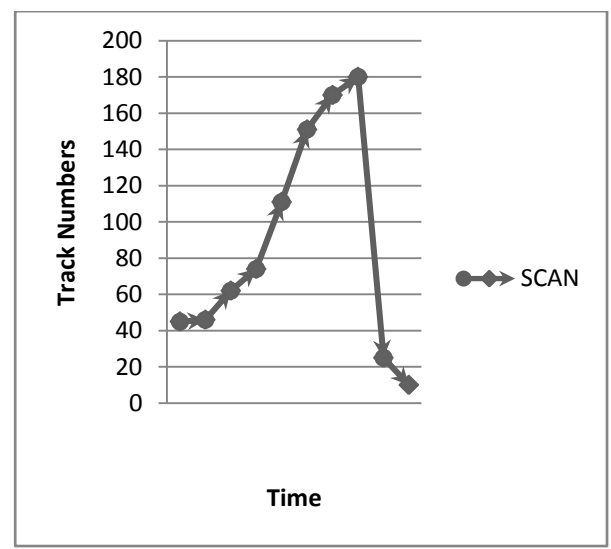

**Fig 5: Representation of SCAN (Case 1)**

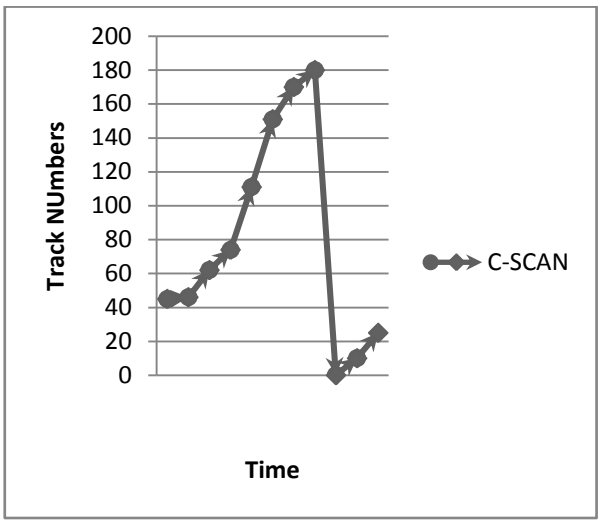

**Fig 6: Representation of C-SCAN (Case 1)**

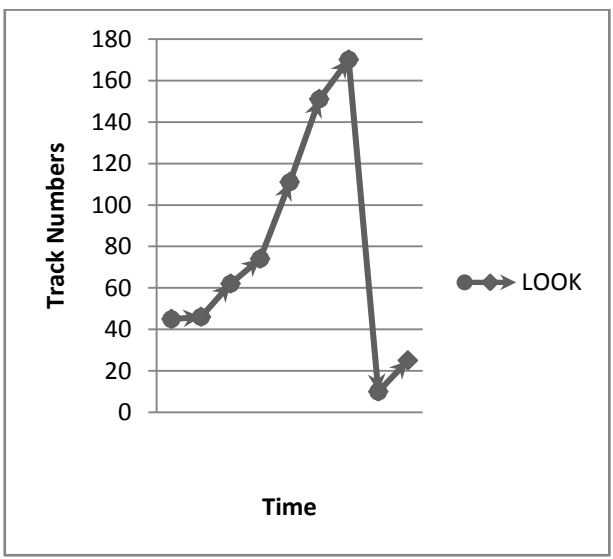

**Fig 7: Representation of LOOK (Case 1)**

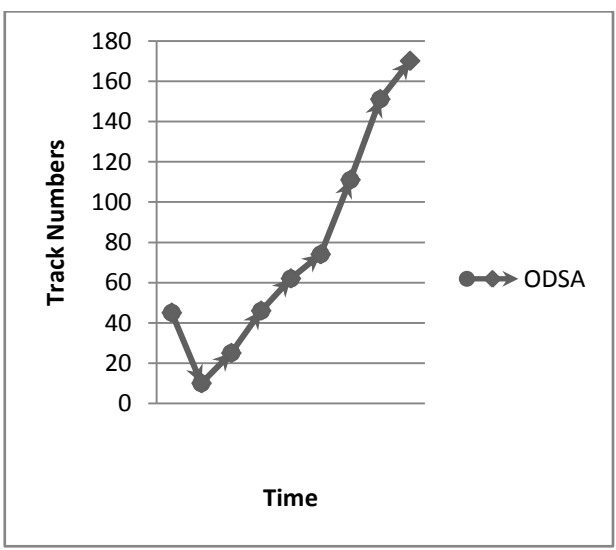

**Fig 8: Representation of ODSA (Case 1)**

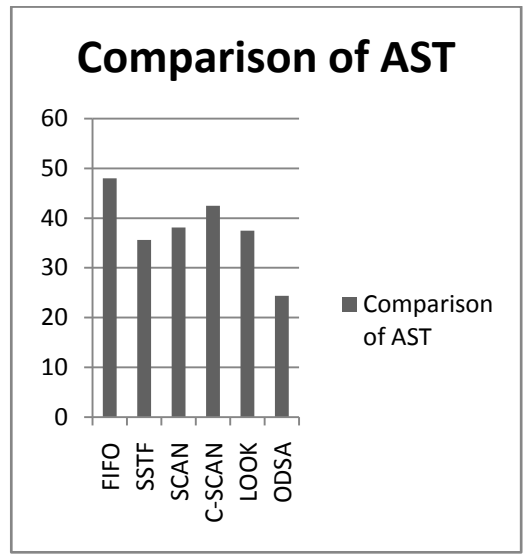

**Fig 9: Comparison of Average Seek Time (Case 1)**

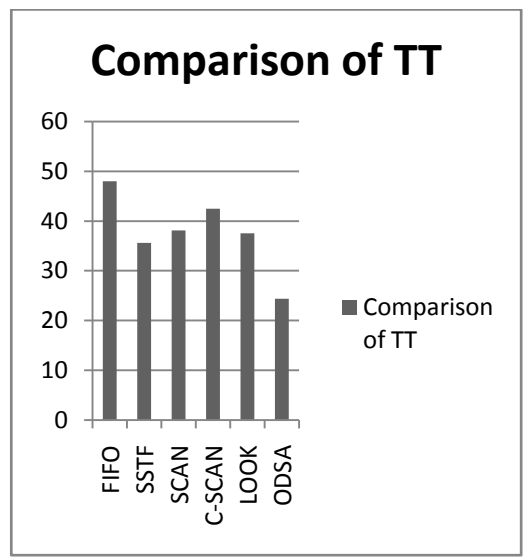

**Fig 10: Comparison of Transfer Time (Case 1)**

**Case 2:** We have taken the following track requests for accessing the tracks as (16, 75, 24, 21, 30, 80, 116 and 63) and the initial position of disk head is at 66. Table 2 shows the comparison of all the algorithms with our proposed algorithm. Figure 11, Figure 12, Figure 13, Figure 14, Figure 15 and Figure 16 shows the representation of FIFO, SSTF, SCAN, C-SCAN, LOOK and ODSA respectively. Figure 17 and Figure 18 shows the comparison of average seek time and transfer time respectively.

**Table 2: Comparison of all algorithms with ODSA**

| Algorithms | Average Seek Time | Transfer Time |
|---|---|---|
| FIFO | 38.875 | 38.88691 |
| SSTF | 19.5 | 19.51191 |
| SCAN | 22.75 | 22.76191 |
| C-SCAN | 43.875 | 43.88691 |
| LOOK | 23.875 | 23.88691 |
| ODSA | 18.75 | 18.76191 |

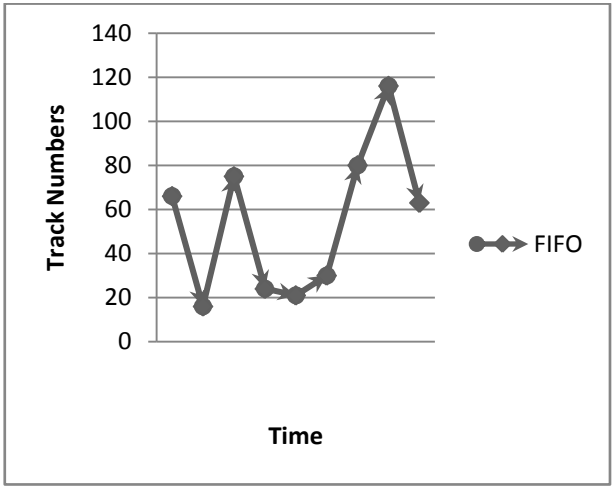

**Fig 11: Representation of FIFO (Case 2)**

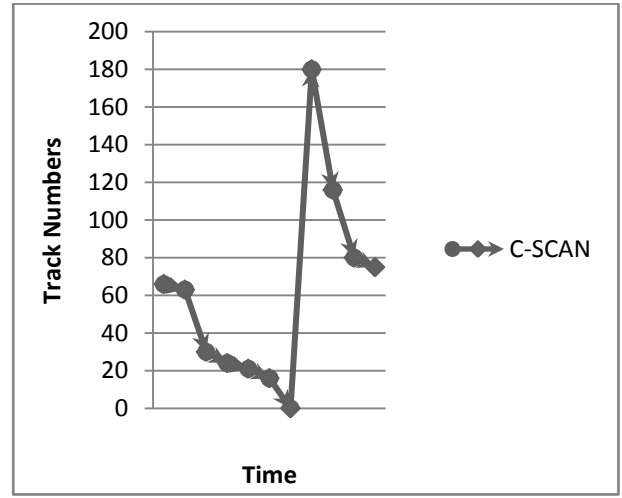

**Fig 14: Representation of C-SCAN (Case 2)**

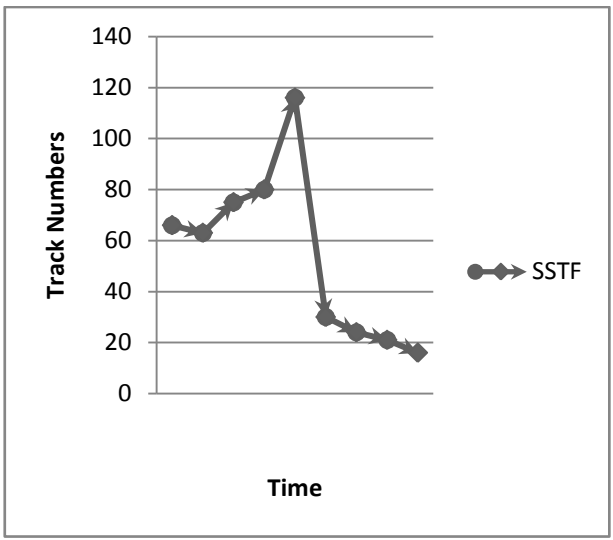

**Fig 12: Representation of SSTF (Case 2)**

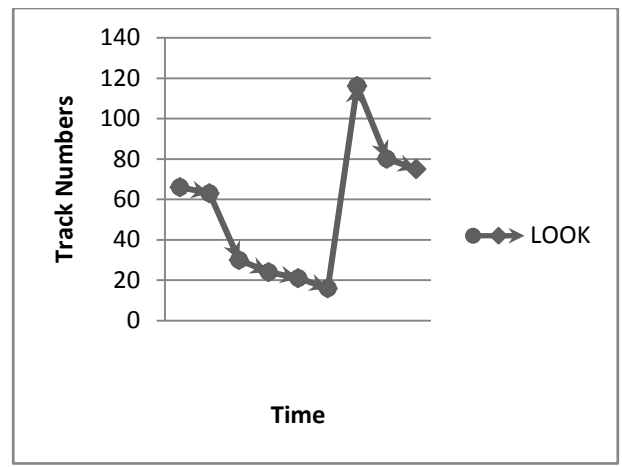

**Fig 15: Representation of LOOK (Case 2)**

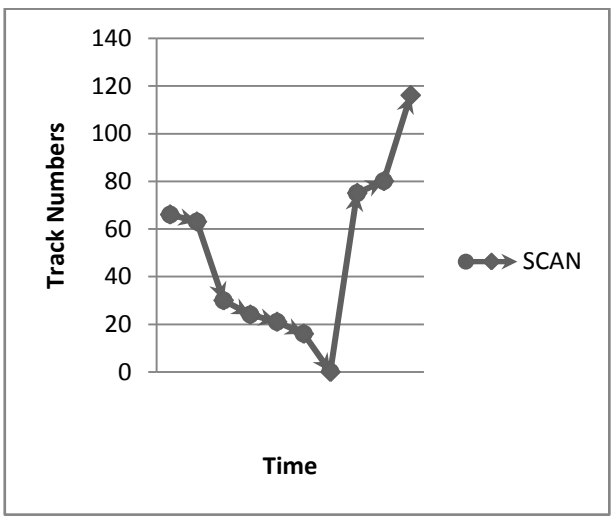

**Fig 13: Representation of SCAN (Case 2)**

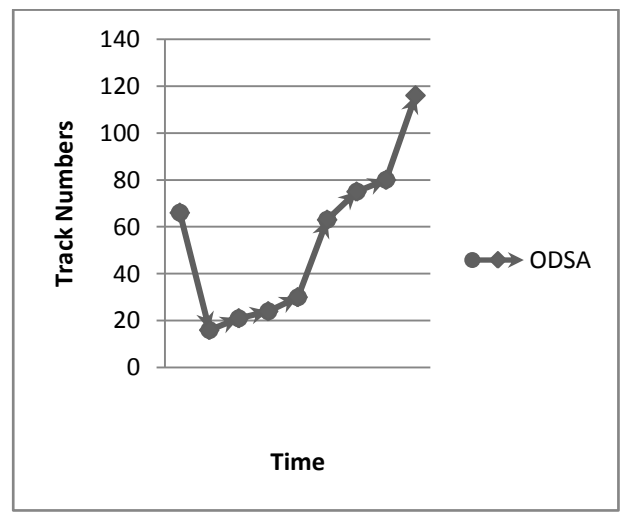

**Fig 16: Representation of ODSA (Case 2)**

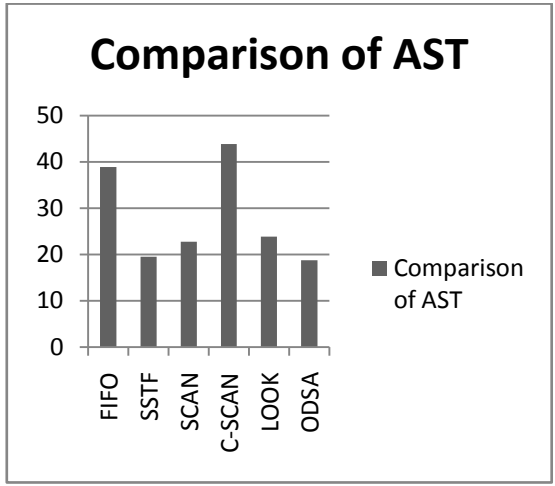

**Fig 17: Comparison of Average Seek Time (Case 2)**

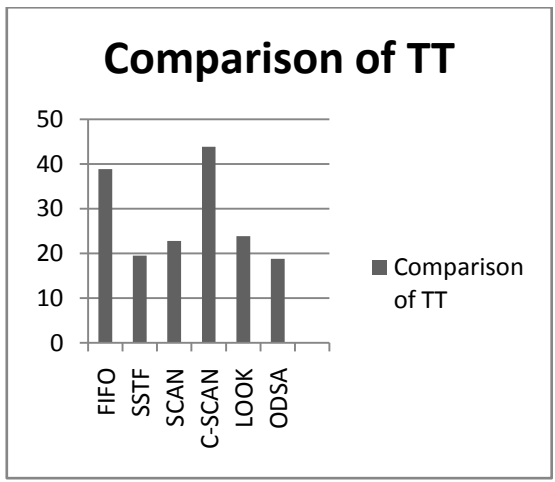

**Fig 18: Comparison of Transfer Time (Case 2)**

**Case 3:** We have taken the following track requests for accessing the tracks as (25, 33, 54, 64, 40, 90, 110 and 160) and the initial disk position is at 125. Table 3 shows the comparison of all the algorithms with our proposed algorithm. Figure 19, Figure 20, Figure 21, Figure 22, Figure 23 and Figure 24 shows the representation of FIFO, SSTF, SCAN, C-SCAN, LOOK and ODSA respectively. Figure 25 and Figure 26 shows the comparison of average seek time and transfer time respectively.

**Table 1: Comparison of all algorithms with ODSA**

| Algorithms | Average Seek Time | Transfer Time |
|---|---|---|
| FIFO | 35.375 | 35.38691 |
| SSTF | 29.375 | 29.38691 |
| SCAN | 35.625 | 35.63691 |
| C-SCAN | 40.625 | 40.63691 |
| LOOK | 29.375 | 29.38691 |
| ODSA | 21.25 | 21.26191 |

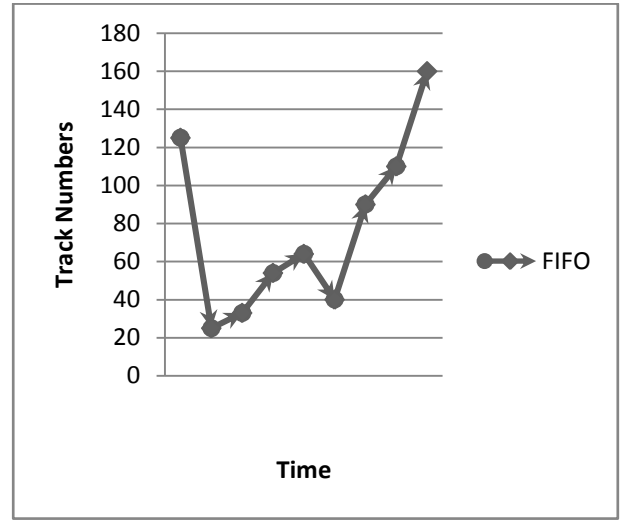

**Fig 19: Representation of FIFO (Case 3)**

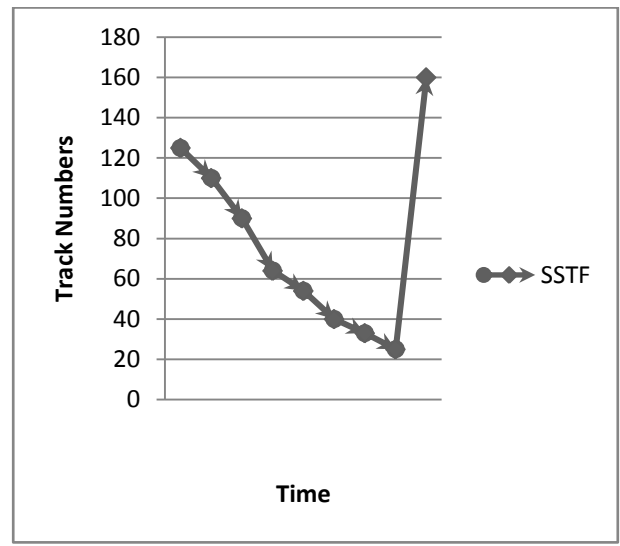

**Fig 20: Representation of SSTF (Case 3)**

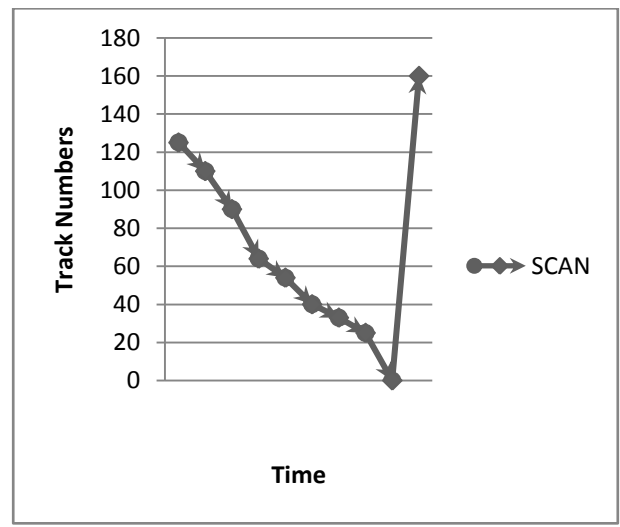

**Fig 21: Representation of SCAN (Case 3)**

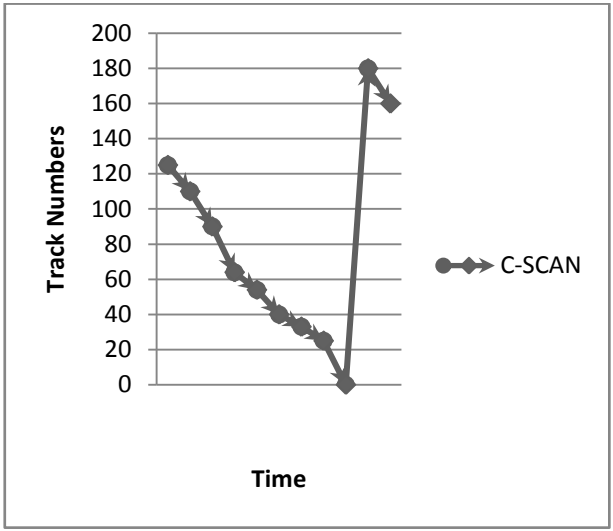

**Fig 22: Representation of C-SCAN (Case 3)**

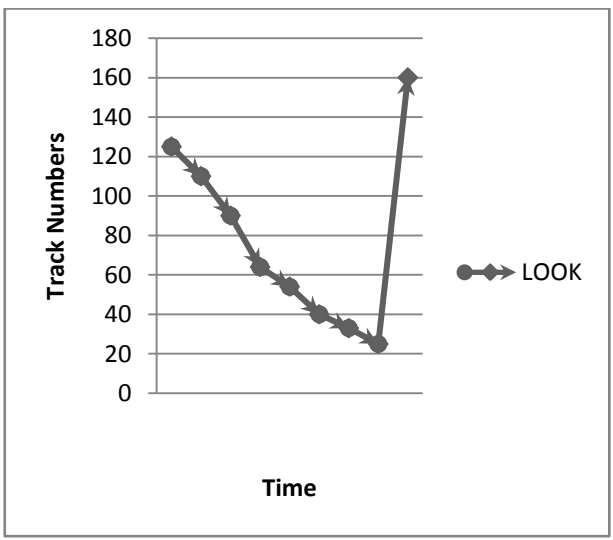

**Fig23: Representation of LOOK (Case 3)**

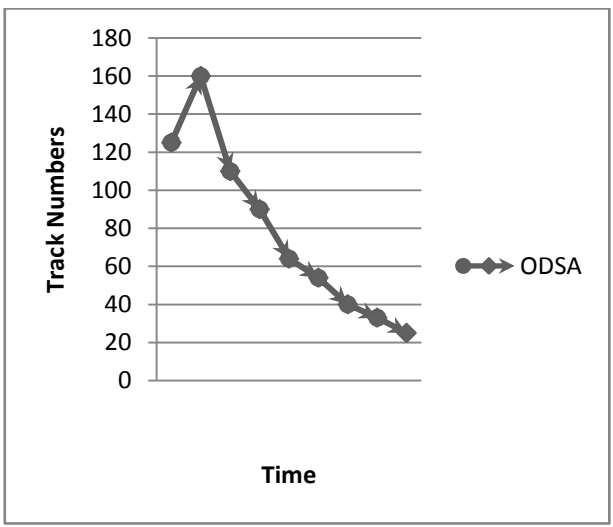

**Fig 24: Representation of ODSA (Case 3)**

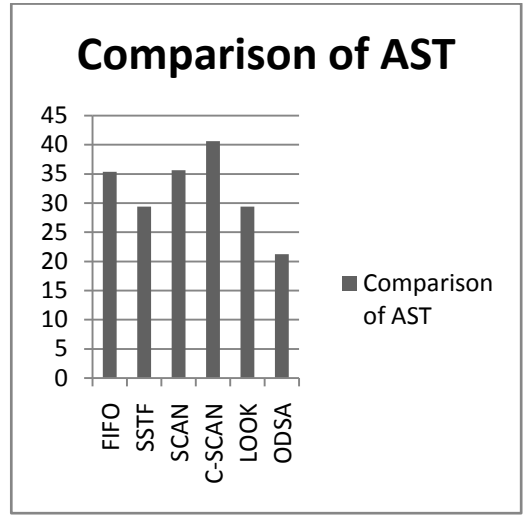

**Fig 25: Comparison of Average Seek Time (Case 3)**

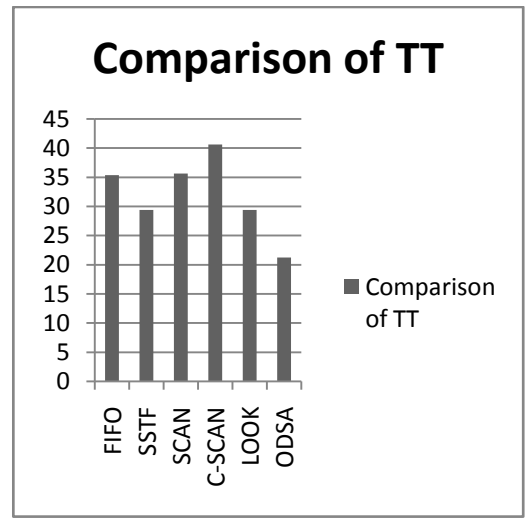

**Fig 26: Comparison of Transfer Time (Case 3)**

## 4. CONCLUSION

The proposed ODSA algorithm shows better performance than other disk scheduling algorithms (FIFO, SSTF, SCAN, C-SCAN and LOOK). The average seek time and transfer time has been improvised by this algorithm which increases the efficiency of the disk performance. In future we can implement this ODSA algorithm in real time systems.

## 5. REFERENCES

[1] H. M. Deitel, "Operating Systems", 2nd Edn., Pearson Education Pte. Ltd., 2002, ISBN 81-7808-035-4.

[2] W. Stallings, "Operating Systems", 4th Edn., Pearson Education Pte. Ltd., 2007, ISBN 81-7808-503-8.

[3] A. Silberschatz, P. B. Galvin and G. Gagne, "Operating System Principles", 7th Edn., John Wiley and Sons, 2008, ISBN 978-81-265-0962-1.

[4] Z. Dimitrijevic, R. Rangaswami and E. Y. Chang, "Support for Preemptive Disk Scheduling", IEEE Transactions on computers, Vol. 54, No. 10, Oct 2005.